\documentclass{article}
\usepackage{amsmath,amssymb,color,url}
\usepackage[pdftex,hiresbb]{graphicx}
\usepackage{type1cm}
\usepackage{subcaption}
\usepackage{geometry}
\allowdisplaybreaks
\newtheorem{thm}{Theorem}[section]

\newtheorem{lem}[thm]{Lemma}
\newtheorem{prop}[thm]{Proposition}
\newtheorem{defn}[thm]{Definition}
\newtheorem{rem}[thm]{Remark}

\newtheorem{ex}[thm]{Example}
\def\proof {\noindent{\it Proof.}$\quad$$\quad$}
\def\fin   {\hfill{$\Box$}}
\def\l     {\left}
\def\r     {\right}
\def\la    {\langle}
\def\ra    {\rangle}

\def\erfc  {\mathop{\mathrm{erfc}}\nolimits}

\def\Re    {\mathop{\mathrm{Re}}\nolimits}
\def\Im    {\mathop{\mathrm{Im}}\nolimits}
\def\arctan{\mathop{\mathrm{arctan}}\nolimits}
\def\sgn   {\mathop{\mathrm{sgn}}\nolimits}
\def\calB  {{\cal B}}

\def\calF  {{\cal F}}

\def\calV  {{\cal V}}
\def\bbC   {{\mathbb C}}
\def\bbD   {{\mathbb D}}
\def\bbE   {{\mathbb E}}
\def\bbF   {{\mathbb F}}

\def\bbN   {{\mathbb N}}
\def\bbP   {{\mathbb P}}

\def\bbR   {{\mathbb R}}
\def\ve    {\varepsilon}
\def\vp    {\varphi}
\def\vt    {\vartheta}

\def\tN    {\widetilde{N}}
\def\theequation{\thesection.\arabic{equation}}
\begin{document}
\title{Pricing and hedging of VIX options for Barndorff-Nielsen and Shephard models}
\author{Takuji Arai\footnote{Department of Economics, Keio University, e-mail:arai@econ.keio.ac.jp}}

\maketitle

\begin{abstract}
The VIX call options for the Barndorff-Nielsen and Shephard models will be discussed.
Derivatives written on the VIX, which is the most popular volatility measurement, have been traded actively very much.
In this paper, we give representations of the VIX call option price for the Barndorff-Nielsen and Shephard models: non-Gaussian Ornstein--Uhlenbeck type stochastic volatility models.
Moreover, we provide representations of the locally risk-minimizing strategy constructed by a combination of the underlying riskless and risky assets.
Remark that the representations obtained in this paper are efficient to develop a numerical method using the fast Fourier transform.
Thus, numerical experiments will be implemented in the last section of this paper.
\end{abstract}


%
%
\setcounter{equation}{0}
\section{Introduction}
Our main objectives are to provide numerically efficient representations of the prices and the locally risk-minimizing (LRM) strategies for the VIX call options
for the Barndorff-Nielsen and Shephard (BNS) models, and implement numerical experiments using the fast Fourier transform (FFT).

The BNS models are non-Gaussian Ornstein--Uhlenbeck (OU)-type stochastic volatility models undertaken by Barndorff--Nielsen and Shephard \cite{BNS1}, \cite{BNS2}.
More precisely, we consider throughout a financial market model composed of one riskless asset with interest rate $r\geq0$ and one risky asset whose price at time $t\geq0$,
denoted by $S_t$, is expressed as
\begin{equation}\label{eq-S}
S_t=S_0\exp\l\{(r+\mu)t-\frac{1}{2}\int_0^t\sigma_s^2ds+\int_0^t\sigma_sdW_s+\rho H_{\lambda t}\r\},
\end{equation}
where $S_0>0$, $\mu\in\bbR$, $\rho\leq0$ and $\lambda>0$.
Here, $W$ is a $1$-dimensional Brownian motion, and $H$ is a subordinator without drift.
The squared volatility process $\sigma^2$ is given by an OU process driven by $H$, that is, the solution to the following equation:
\[
d\sigma_t^2=-\lambda\sigma_t^2dt+dH_{\lambda t}
\]
with $\sigma_0^2>0$.
In this paper, we take $\mu\in\bbR$ so that the discounted asset price process $e^{-rt}S_t$, denoted by $\widehat{S}_t$, becomes a martingale.
Thus, for any option $X$ matured at time $T>0$, its price at time $t\in[0,T]$ is given by $e^{-r(T-t)}\bbE[X|\calF_t]$, where $\{\calF_t\}_{t\geq0}$ is a filtration.
On the other hand, since our underlying market is incomplete, there is no perfect hedge in general.
Instead, we consider an alternative hedging strategy, which is not perfect, but optimal in some sense.
Actually, many such hedging strategies for incomplete market models have been suggested.
Among them, we focus on the LRM strategy, which is a very well-known quadratic hedging method.
In particular, its theoretical aspects have been well developed, but little is known about its explicit representations.
Meanwhile, Arai et al. \cite{AIS-BNS} gave a representation of the LRM strategies for call options for BNS models using Malliavin calculus for L\'evy processes,
and illustrated an FFT-based numerical method.

Now, the VIX is the most popular volatility measurement launched by the Chicago Board Options Exchange (CBOE) \cite{CBOE}.
More precisely, it is defined as the square root of the expected value of integrated variance of the S\&P 500 index over the next 30 business days.
In this paper, the VIX at time $t$, denoted by $\calV_t$, is defined as the integrated variance over the time interval $[t, t+\tau]$,
where $\tau>0$ is the fixed observation period.
As seen in Section 2, the mathematical definition of the square of $\calV_t$ is naturally given as
\begin{equation}\label{def-VIX}
\calV^2_t:=-\frac{2}{\tau}\bbE\l[\log\widehat{S}_{t+\tau}-\log\widehat{S}_t|\calF_t\r].
\end{equation}
In addition, the right-hand side of (\ref{def-VIX}) is rewritten as
\begin{equation}\label{def-VIX2}
\calV^2_t=\frac{1}{\tau}\bbE\l[\int_t^{t+\tau}\sigma^2_sds\Big|\calF_t\r] - 2\int_0^\infty\l(1+\rho x-e^{\rho x}\r)\nu(dx).
\end{equation}
Remark that the second term is due to the jump component of (\ref{eq-S}).
It is well-known that changes of the VIX are negatively correlated to changes in asset prices, but it is not directly investable.
Thus, trades of derivatives written on the VIX are inevitable in order to reduce risks caused by changes of volatility.
Actually, such derivatives have been traded actively very much, and there are much literature on this topic.
In this paper, we focus on the European-type call options written on the VIX, of which  payoff is described as $(\calV_T-K)^+$, where $T>0$ is maturity and $K>0$ is strike price;
and provide representations of their prices and LRM strategies for the BNS models by extending results in \cite{AIS-BNS},
where the LRM strategies discussed in this paper are given by a combination of the underlying riskless and risky assets.
In particular, our representations obtained in this paper are efficient to develop an FFT-based numerical method.

Pricing and hedging problems of derivatives on the VIX or volatilities for jump type models have already been studied by a number of researchers
(\cite{BNP}, \cite{DK}, \cite{GM}, \cite{JMM}, \cite{LLZ}, \cite{Sepp1}, \cite{Sepp2}, \cite{SW} and so forth).
Among them, Lian and Zhu \cite{LZ} derived a pricing formula of the VIX call options for the so-called SVJJ models,
in which the asset price process $S^S$ is given as the solution to the following stochastic differential equation (SDE):
\begin{equation}\label{SVJJ}
dS^S_t=S^S_{t-}\l\{\mu^Sdt+\sigma^S_tdW^S_t+dZ^S_t\r\},
\end{equation}
where $\mu^S\in\bbR$, $W^S$ is a $1$-dimensional Brownian motion, $Z^S$ is a compound Poisson process.
Note that their formula was obtained as a correction of \cite{LC}. In addition, \cite{CILZ} also pointed out errors of \cite{LC}.
Here, $\sigma^S$ in (\ref{SVJJ}) is a volatility process given from the solution to the following SDE:
\[
d(\sigma^S_t)^2=\kappa^S(\theta^S-(\sigma^S_t)^2)dt+v^S\sigma^S_tdW^\sigma_t+dZ^\sigma_t,
\]
where $\kappa^S$, $\theta^S$ and $v^S$ are constants satisfying the Feller condition
\[
2\kappa^S\theta^S>(v^S)^2,
\]
$W^\sigma$ is a $1$-dimensional Brownian motion correlated to $W^S$,
and $Z^\sigma$ is a compound Poisson process generated by the same Poisson process as $Z^S$, that is, jumps of $Z^S$ and $Z^\sigma$ happen simultaneously.
Remark that this model framework does not include the BNS models.

As another literature, Barletta and Nicolato \cite{BN} also considered the same model framework as \cite{LZ},
and derived a closed-form pricing formula using approximations via orthogonal expansions.
Kallsen et al. \cite{KMV} studied pricing of options written on the quadratic variation of a given asset price process for affine stochastic volatility models with jumps.
In particular, they illustrated numerical experiments for BNS models.
Benth et al. \cite{BGK} obtained a valuation formula for conditional expectations of powers of the realized volatility
\[
\sigma_R:=\sqrt{\frac{1}{T}\int_0^T\sigma^2_sds}
\]
for the BNS models without jumps, that is, the case where $\rho=0$.
Note that the realized volatility $\sigma_R$ used in \cite{BGK} is different from the VIX $\calV$ defined in (\ref{def-VIX2}).
Moreover, Habtemicael and SenGupta \cite{HS} and Issaka and SenGupta \cite{IS} studied the variance swap $\sigma_R^2-K_R$ and the volatility swap $\sigma_R-K_R$
for the BNS models with jumps, where $K_R$ is delivery price.
In particular, \cite{IS} derived a partial integro-differential equation describing the price dynamics of the variance swaps, and a Ve\v{c}e\v{r}-type formula.
To our best knowledge, any representations of the prices and the LRM strategies of the VIX call options for the BNS models have not been provided.
In particular, no one has discussed the LRM strategies for the VIX options for jump-type stochastic volatility models.

The rest of the paper is organized as follows:
In Section 2, we give model description, and discuss the VIX.
Our main results, that is, representations of the prices and the LRM strategies of the VIX call options are given in Sections 3 and 4, respectively.
Section 5 is devoted to numerical results.

%
%
\setcounter{equation}{0}
\section{Preliminaries}
\subsection{Model description}
We consider, throughout this paper, a financial market being composed of one riskless asset with interest rate $r\geq0$
and one risky asset whose price dynamics is described by (\ref{eq-S}).
Note that the risky asset price process $S$ is also given as the solution to the following SDE:
\begin{equation}\label{SDE0}
\frac{dS_t}{S_{t-}}=\l(r+\mu+\int_0^\infty(e^{\rho x}-1)\nu(dx)\r)dt+\sigma_tdW_t+\int_0^\infty(e^{\rho x}-1)\tN(dt,dx).
\end{equation}
Here $N$ denotes the Poisson random measure of the subordinator $H_{\lambda t}$, that is, 
\[
H_{\lambda t}=\int_0^\infty xN([0,t],dx)
\]
holds for $t\geq0$, $\nu$ is the L\'evy measure of $H_{\lambda t}$; and $\tN$ is the compensated version of $N$, which is represented as
\[
\tN(dt,dx)=N(dt,dx)-\nu(dx)dt.
\]
Remark that the last term $\rho H_{\lambda t}$ in (\ref{eq-S}) accounts for the leverage effect,
which is a stylized fact such that the asset price declines at the moment when volatility increases.
In this paper, we treat only the case where the discounted asset price process $\widehat{S}_t$($:=e^{-rt}S_t$) becomes a martingale.
In other words, $\mu$ is assumed to be given as $\int_0^\infty(1-e^{\rho x})\nu(dx)$.
Thus, (\ref{SDE0}) implies that the dynamics of $\widehat{S}$ is given by
\[
\frac{d\widehat{S}_t}{\widehat{S}_{t-}}=\sigma_tdW_t+\int_0^\infty(e^{\rho x}-1)\tN(dt,dx).
\]

\begin{rem}\label{rem-1}
We shall use Malliavin calculus based on the canonical L\'evy space, undertaken by Sol\'e et al. \cite{S07}.
Thus, the underlying probability space $(\Omega, \calF, \bbP)$ is supposed to be given as the product space $(\Omega_W\times\Omega_J, \calF_W\times\calF_J, \bbP_W\times\bbP_J)$,
where $(\Omega_W, \calF_W, \bbP_W)$ and $(\Omega_J, \calF_J, \bbP_J)$ are a one-dimensional Wiener space and
the canonical L\'evy space for the pure jump L\'evy process $H_{\lambda t}$, respectively.
A filtration $\bbF=\{\calF_t\}_{t\geq0}$ denotes the canonical filtration completed for $\bbP$.
Although the results obtained in this paper are basically not depending on the structure of the underlying probability space,
we choose the canonical L\'evy space framework in order to simplify mathematical description and discussion.
For example, it is possible to use results on the canonical L\'evy space introduced in Arai and Suzuki \cite{AS}, Delong and Imkeller \cite{DI} and Suzuki \cite{Suz}.
\end{rem}

As seen in Introduction, the volatility process $\sigma$ in (\ref{SDE0}) is a square root of an OU process driven by the subordinator $H_{\lambda t}$.
Now, we introduce two important examples of the squared volatility process $\sigma^2$ appeared in BNS models.
For more details on this topic, see also Schoutens \cite{Scho} and Nicolato and Venardos \cite{NV}.
\begin{enumerate}
\item The first one is the case where $\sigma^2$ follows an IG-OU process. The corresponding L\'evy measure $\nu$ is given by
      \[
      \nu(dx)=\frac{\lambda a}{2\sqrt{2\pi}}x^{-\frac{3}{2}}(1+b^2x)e^{-\frac{1}{2}b^2x}{\bf 1}_{(0,\infty)}(x)dx
      \]
      where $a>0$ and $b>0$.
      Note that this is a representative example of BNS models with infinite active jumps, that is, $\nu((0,\infty))=\infty$.
      In this case, the invariant distribution of $\sigma^2$ follows an inverse-Gaussian distribution with parameters $a>0$ and $b>0$.
\item The second example is the gamma-OU case. In this case, $\nu$ is described as
      \[
      \nu(dx)=\lambda abe^{-bx}{\bf 1}_{(0,\infty)}(x)dx,
      \]
      and the invariant distribution of $\sigma^2$ is given by a gamma distribution with parameters $a>0$ and $b>0$.
\end{enumerate}

\subsection{VIX}
In this subsection, we discuss the reason why the VIX is defined as in (\ref{def-VIX}), and show that (\ref{def-VIX2}) holds for the BNS models.
To this end, we firstly consider a continuous-type stochastic volatility model in which the discounted asset price process $\widehat{S}^C$ is given as
\[
d\widehat{S}^C_t=\widehat{S}^C_t\sigma^C_tdW_t.
\]
Note that we do not need to specify the volatility process $\sigma^C_t$.
The square of the VIX for this model, denoted by $(\calV^C)^2$, is naturally defined as
\[
(\calV^C_t)^2=\frac{1}{\tau}\bbE\l[\int_t^{t+\tau}(\sigma^C_s)^2ds\Big|\calF_t\r]
\]
for $t\in[0,T]$, where $\tau>0$ is the observation period.
By simple calculation, we have
\begin{equation}\label{eq-modelfree}
\bbE\l[\int_t^{t+\tau}(\sigma^C_s)^2ds\Big|\calF_t\r]=-2\bbE\l[\log\widehat{S}^C_{t+\tau}-\log\widehat{S}^C_t|\calF_t\r].
\end{equation}
On the other hand, due to the jump component, the integrated variance over $[t,t+\tau]$ for the BNS models is different from $\bbE\l[\int_t^{t+\tau}\sigma^2_sds\Big|\calF_t\r]$.
Thus, taking account of (\ref{eq-modelfree}), we define the square of the VIX for the BNS models as in (\ref{def-VIX}).
In order to treat the VIX $\calV$ on the time interval $[0,T]$, the processes $S$ and $\sigma^2$ should be defined on the extended time interval $[0,T+\tau]$,
where $T>0$ is the maturity of the option to be priced and hedged.

In order to make sure of (\ref{def-VIX2}), we calculate $\calV^2_t$ as follows:
\begin{align}\label{def-VIX3}
\calV^2_t
&:= -\frac{2}{\tau}\bbE\l[\log\widehat{S}_{t+\tau}-\log\widehat{S}_t|\calF_t\r] \nonumber \\
&=  -\frac{2}{\tau}\bbE\Bigg[\int_t^{t+\tau}\l(\int_0^\infty\l(1-e^{\rho x}\r)\nu(dx)-\frac{1}{2}\sigma^2_s\r)ds \nonumber \\
&   \hspace{5mm}+\int_t^{t+\tau}\sigma_sdW_s+\int_t^{t+\tau}\int_0^\infty\rho xN(ds,dx)\Big|\calF_t\Bigg] \nonumber \\
&=  \frac{1}{\tau}\bbE\l[\int_t^{t+\tau}\sigma^2_sds\Big|\calF_t\r] - 2\int_0^\infty\l(1+\rho x-e^{\rho x}\r)\nu(dx)
\end{align}
for $t\in[0,T]$.
Furthermore, the first term of the right-hand side of (\ref{def-VIX3}) is given as
\begin{align*}
\frac{1}{\tau}\bbE\l[\int_t^{t+\tau}\sigma^2_sds|\calF_t\r]
&= \frac{1}{\tau}\bbE\l[\calB(\tau)\sigma^2_t+\int_t^{t+\tau}\int_0^\infty\calB(t+\tau-s)xN(ds,dx)\Big|\calF_t\r] \\
&= \frac{\calB(\tau)}{\tau}\sigma^2_t+\frac{1}{\tau}\int_t^{t+\tau}\calB(t+\tau-s)ds\int_0^\infty x\nu(dx) \\
&= \frac{\calB(\tau)}{\tau}\sigma^2_t+\frac{1}{\lambda}\l(1-\frac{\calB(\tau)}{\tau}\r)\int_0^\infty x\nu(dx)
\end{align*}
by (2.5) of \cite{NV}, where
\[
\calB(t):=\frac{1-e^{-\lambda t}}{\lambda}
\]
for $t\geq0$.
As a result, we have and denote
\begin{align*}
\calV^2_t &=  \frac{\calB(\tau)}{\tau}\sigma^2_t+\frac{1}{\lambda}\l(1-\frac{\calB(\tau)}{\tau}\r)\int_0^\infty x\nu(dx) - 2\int_0^\infty\l(1+\rho x-e^{\rho x}\r)\nu(dx) \\
          &=: B_\calV\sigma^2_t+C_\calV,
\end{align*}
where $B_{\calV}$ and $C_{\calV}$ are positive constants.
Thus, we can describe $\calV_t$ as
\begin{equation}\label{eq-VIX}
\calV_t = \sqrt{B_\calV\sigma^2_t+C_\calV}.
\end{equation}

%
%
\setcounter{equation}{0}
\section{Pricing}
The aim of this section is to provide two representations of the prices of the VIX call options for the BNS models.
Note that our representations are efficient to develop an FFT-based numerical scheme.
Firstly, we give an integral expression under an integrable condition on the characteristic function of $\sigma^2_T$.
Note that this condition is satisfied in the IG-OU case, but not in the gamma-OU case.
Thus, we suggest alternatively an approximate method in order to treat the gamma-OU case.

Consider the VIX call option matured at time $T>0$ with strike price $K>0$.
Then, its payoff is described as  $(\calV_T-K)^+$; and its price at time $t$, denoted by $P_t$, is given as follows:
\[
P_t:= e^{-r(T-t)}\bbE[(\calV_T-K)^+|\calF_t].
\]
In addition, we define the Fourier transform of the payoff function of the VIX call option as
\[
\widehat{g}(v,\alpha;K):=\int_0^\infty(\sqrt{B_\calV x+C_\calV}-K)^+e^{(iv-\alpha)x}dx
\]
for $v\in\bbR$ and $\alpha>0$.
Note that, since $\sigma^2_T$ is positive, $\widehat{g}$ is defined as an integration on $[0,\infty)$ instead of $\bbR$;
and it is enough to treat only the case where $K\geq\sqrt{C_\calV}$.
A concrete expression of $\widehat{g}$ is given as follows:

\begin{lem}\label{lem-g}
For any $K\geq\sqrt{C_\calV}$, $v\in\bbR$ and $\alpha>0$, we have
\[
\widehat{g}(v,\alpha;K) = \exp\l\{-\frac{(iv-\alpha)C_\calV}{B_\calV}\r\}\frac{\sqrt{B_\calV\pi}}{2(-iv+\alpha)^{\frac{3}{2}}}\erfc\l(K\sqrt{\frac{-iv+\alpha}{B_\calV}}\r),
\]
where
\[
\erfc(x):=\frac{2}{\sqrt{\pi}}\int_x^\infty e^{-t^2}dt.
\]
\end{lem}

\proof
By (9) of \cite{LZ}, we have
\begin{align*}
\lefteqn{\int_0^\infty(\sqrt{B_\calV x+C_\calV}-K)^+e^{(iv-\alpha)x}dx} \\
&= \exp\l\{-\frac{(iv-\alpha)C_\calV}{B_\calV}\r\}\int_{C_\calV}^\infty(\sqrt{y}-K)^+e^{\frac{(iv-\alpha)y}{B_\calV}}\frac{dy}{B_\calV} \\
&= \exp\l\{-\frac{(iv-\alpha)C_\calV}{B_\calV}\r\}\int_0^\infty(\sqrt{y}-K)^+e^{\frac{(iv-\alpha)y}{B_\calV}}\frac{dy}{B_\calV} \\
&= \exp\l\{-\frac{(iv-\alpha)C_\calV}{B_\calV}\r\}\frac{\sqrt{B_\calV\pi}}{2(-iv+\alpha)^{\frac{3}{2}}}\erfc\l(K\sqrt{\frac{-iv+\alpha}{B_\calV}}\r),
\end{align*}
from which Lemma \ref{lem-g} follows.
\fin

In order to give an expression of $P_t$, we need to define the conditional characteristic function of $\sigma^2_T$ given $\sigma^2_t$ as
\[
\phi_{T|t}(\zeta):=\bbE[\exp\{i\zeta\sigma^2_T\}|\sigma^2_t]
\]
for $\zeta\in\bbC$.
Lemma 2.1 of \cite{NV} implies that, denoting
\[
\kappa(u):=\int_0^\infty(e^{ux}-1)\nu(dx),
\]
we have
\begin{align}\label{eq-phi}
\phi_{T|t}(\zeta)
&= \bbE\l[\exp\l\{i\zeta e^{-\lambda(T-t)}\sigma_t^2+i\zeta\int_t^Te^{-\lambda(T-s)}dH_{\lambda s}\r\}\Big|\sigma^2_t\r] \nonumber \\
&= \exp\l\{i\zeta e^{-\lambda(T-t)}\sigma_t^2\r\}\exp\l\{\int_t^T\kappa\l(i\zeta e^{-\lambda(T-s)}\r)ds\r\}
\end{align}
for any $\zeta\in\bbC$ with $\Im(\zeta)>-\widehat{u}$, where
\[
\widehat{u}:=\sup\{u\in\bbR | \kappa(u)<\infty\}\geq 0.
\]
Now, $P_t$ has the following integration expression:

\begin{prop}[Proposition 2 of Tankov \cite{T}]\label{prop-Tankov}
Suppose that $\widehat{u}>0$ and
\begin{equation}\label{eq-phicond}
\int_\bbR\frac{|\phi_{T|t}(v-i\alpha)|}{1+|v|}dv<\infty
\end{equation}
for any $t\in[0,T]$ and $\alpha\in(0,\widehat{u})$.
We have then
\begin{equation}\label{eq-Pt}
P_t =  \frac{e^{-r(T-t)}}{2\pi}\int_{-\infty}^\infty\widehat{g}(v,\alpha;K)\phi_{T|t}(-v-i\alpha)dv
\end{equation}
for any $t\in[0,T]$, $\alpha\in(0,\widehat{u})$ and $K\geq\sqrt{C_\calV}$.
Note that the right-hand side of (\ref{eq-Pt}) is independent of the choice of $\alpha$.
\end{prop}

\begin{rem}
The above expression (\ref{eq-Pt}) has been already introduced in Proposition 2 of \cite{LZ} for the SVJJ models, but the BNS models are not included.
\end{rem}

\begin{rem}
As another important derivative written on the VIX, the VIX futures has been traded actively.
Its value at time $t$ is denoted by
\[
F^\calV_t:=\bbE[\calV_T|\calF_t],
\]
which is corresponding to the price of the VIX call option with strike price $0$ when the interest rate $r$ is also $0$.
Thus, (\ref{eq-Pt}) implies that
\[
F^\calV_t = \frac{1}{2\pi}\int_{-\infty}^\infty\widehat{g}(v,\alpha;0)\phi_{T|t}(-v-i\alpha)dv
\]
holds under all the conditions of Proposition \ref{prop-Tankov}.
\end{rem}

We show that the IG-OU case introduced in Subsection 2.1 satisfies all the conditions of Proposition \ref{prop-Tankov}, that is,
the VIX option prices for the IG-OU case are described as (\ref{eq-Pt}).

\begin{ex}\label{ex-IG-OU}
Firstly, (2.8) of \cite{NV} implies that
\[
\kappa(u)=\int_0^\infty(e^{ux}-1)\nu(dx)=\lambda au(b^2-2u)^{-\frac{1}{2}}
\]
for $u<\frac{b^2}{2}$, which means $\widehat{u}=\frac{b^2}{2}>0$.
Next, we show that the condition (\ref{eq-phicond}) is satisfied for any $a$, $b>0$.
To this end, we calculate $\int_t^T\kappa\l(i\zeta e^{-\lambda(T-s)}\r)ds$ for $\zeta\in\bbC$ with $\Im(\zeta)>-\frac{b^2}{2}$ as follows:
\begin{align}
\lefteqn{\int_t^T\kappa\l(i\zeta e^{-\lambda(T-s)}\r)ds} \nonumber \\
&= \int_t^T\frac{\lambda ai\zeta e^{-\lambda(T-s)}}{\sqrt{b^2-2i\zeta e^{-\lambda(T-s)}}}ds
   = \int_{e^{-\lambda(T-t)}}^1\frac{ai\zeta}{\sqrt{b^2-2i\zeta x}}dx \nonumber \\
&= \frac{a\zeta}{\sqrt{2}}\int_{e^{-\lambda(T-t)}}^1\l\{-\sgn(\Re(\zeta))\sqrt{\frac{-A+\sqrt{A^2+B^2}}{A^2+B^2}}+\sqrt{\frac{A+\sqrt{A^2+B^2}}{A^2+B^2}}i\r\}dx
\label{eq-kappa-IG}
\end{align}
where $A:=b^2+2\Im(\zeta)x$ and $B:=2\Re(\zeta)x$.
Taking $v\in\bbR$ and $\alpha\in(0,\widehat{u})$, we substitute $v-i\alpha$ for $\zeta$ to estimate the real part of the integrand of (\ref{eq-kappa-IG}).
We then can find a constant $C>0$ such that
\[
-|v|\sqrt{\frac{-A+\sqrt{A^2+B^2}}{A^2+B^2}}+\alpha\sqrt{\frac{A+\sqrt{A^2+B^2}}{A^2+B^2}}<C\l(-\sqrt{|v|}+\frac{1}{\sqrt{|v|}}\r).
\]
for any $x\in(e^{-\lambda(T-t)},1)$ and any $v\in\bbR$ with sufficient large $|v|$.
Consequently, the IG-OU case always satisfies (\ref{eq-phicond}) from the view of (\ref{eq-phi}).
\end{ex}

For the gamma-OU case, which is another typical framework of the BNS models, the condition (\ref{eq-phicond}) is not satisfied as seen in Example \ref{ex-gamma-OU} below,
that is, the right-hand side of (\ref{eq-Pt}) is not well-defined.
To overcome this difficulty, we develop an approximate method by replacing $\sigma^2_T$ with $\sigma^2_T+\ve(W_T-W_t)$, denoted by $\sigma^2_{T|t}(\ve)$, for sufficient small $\ve>0$.
To this end, we need to consider the VIX of $\sigma^2_{T|t}(\ve)$, instead of $\sigma^2_T$.
Since $\sigma^2_{T|t}(\ve)$ might take negative values, we rewrite the payoff of the VIX call options as
\[
\l(\sqrt{|B_\calV \sigma^2_{T|t}(\ve)+C_\calV|}-K\r){\bf 1}_{\{\sigma^2_{T|t}(\ve)>(K^2-C_\calV)/B_\calV\}},
\]
and define
\[
P^{(\ve)}_t:= e^{-r(T-t)}\bbE\l[\l(\sqrt{|B_\calV\sigma^2_{T|t}(\ve)+C_\calV|}-K\r){\bf 1}_{\{\sigma^2_{T|t}(\ve)>(K^2-C_\calV)/B_\calV\}}|\calF_t\r]
\]
for $\ve>0$.
We have then
\[
\lim_{\ve\to0}P^{(\ve)}_t=P_t,
\]
which means that computing $P^{(\ve)}_t$ for sufficient small $\ve>0$ gives the value of $P_t$ approximately.
Now, we show that $P^{(\ve)}_t$ has the same type integral representation as (\ref{eq-Pt}).

\begin{prop}\label{prop-approx}
Suppose that $\widehat{u}>0$, and, for any $t\in[0,T]$ and $\alpha\in(0,\widehat{u})$, there exists a constant $C>0$ such that
\begin{equation}\label{eq-prop-ass}
|\phi_{T|t}(v-i\alpha)| < C
\end{equation}
for any $v\in\bbR$.
We have then
\begin{equation}\label{eq-Pt-ve}
P^{(\ve)}_t=\frac{e^{-r(T-t)}}{2\pi}\int_{-\infty}^\infty\widehat{g}(v,\alpha;K)\phi_{T|t}(-v-i\alpha)\exp\l\{-\frac{\ve^2(-v-i\alpha)^2(T-t)}{2}\r\}dv
\end{equation}
for $t\in[0,T)$, $\alpha\in(0,\widehat{u})$ and $K\geq\sqrt{C_\calV}$.
Note that the right-hand side of (\ref{eq-Pt-ve}) is independent of the choice of $\alpha$.
\end{prop}

\proof Denoting
\begin{equation}\label{eq-phi-ve}
\phi^{(\ve)}_{T|t}(\zeta):=\bbE[\exp\{i\zeta\sigma^2_{T|t}(\ve)\}|\calF_t]=\bbE[\exp\{i\zeta(\sigma^2_T+\ve(W_T-W_t))\}|\calF_t]
\end{equation}
for $\ve>0$, $t\in[0,T]$ and $\zeta\in\bbC$, we have
\[
\phi^{(\ve)}_{T|t}(\zeta) = \phi_{T|t}(\zeta)\bbE[\exp\{i\zeta\ve W_{T-t}\}] = \phi_{T|t}(\zeta)\exp\l\{-\frac{\zeta^2\ve^2(T-t)}{2}\r\},
\]
which implies that
\[
\int_\bbR\frac{|\phi^{(\ve)}_{T|t}(v-i\alpha)|}{1+|v|}dv<\infty
\]
holds for any $\ve>0$, $t\in[0,T)$ and $\alpha\in(0,\widehat{u})$ by the condition (\ref{eq-prop-ass}).
As a result, we obtain (\ref{eq-Pt-ve}) using Proposition \ref{prop-Tankov}.
\fin

\begin{rem}
Considering
\[
P^{\la N\ra}_t:= \frac{e^{-r(T-t)}}{2\pi}\int_{-N}^N\widehat{g}(v,\alpha;K)\phi_{T|t}(-v-i\alpha)dv
\]
for $N\in\bbN$ instead of $P^{(\ve)}_t$, we expect that it gives a good approximation for $P_t$ by computing $P^{\la N\ra}_t$ for sufficient large $N\in\bbN$,
since integrations on $(-\infty,\infty)$ are computed numerically by truncating the integration interval.
However, $P^{\la N\ra}_t$ never converges to $P_t$ as $N$ tends to $\infty$ for the gamma-OU cases.
\end{rem}

We see that the gamma-OU case satisfies all the conditions of Proposition \ref{prop-approx}.
As a result, we can compute the values of $P^{(\ve)}_t$ numerically using the integral expression (\ref{eq-Pt-ve}),
which approximates the values of $P_t$ when $\ve>0$ is small enough.

\begin{ex}\label{ex-gamma-OU}
Recall that the L\'evy measure $\nu$ in the gamma-OU case is described as
\[
\nu(dx)=\lambda abe^{-bx}{\bf 1}_{(0,\infty)}(x)dx
\]
for $a$, $b>0$.
By (2.10) of \cite{NV},
\[
\kappa(u)=\int_0^\infty(e^{ux}-1)\nu(dx)=\frac{\lambda au}{b-u}
\]
holds for $u<b$, and $\widehat{u}=b$, which is positive.
For $\zeta\in\bbC$ with $\Im(\zeta)>-b$, we have
\begin{align}
\lefteqn{\int_t^T\kappa\l(i\zeta e^{-\lambda(T-s)}\r)ds} \nonumber \\
&= \int_t^T\frac{\lambda ai\zeta e^{-\lambda(T-s)}}{b-i\zeta e^{-\lambda(T-s)}}ds
   = \int_{e^{-\lambda(T-t)}}^1\frac{ai\zeta}{b-i\zeta x}dx = a\zeta\int_{e^{-\lambda(T-t)}}^1\frac{bi-\overline{\zeta}x}{|b-i\zeta x|^2}dx \nonumber \\
&= a\int_{e^{-\lambda(T-t)}}^1\frac{b\zeta i-\zeta\overline{\zeta}x}{\zeta\overline{\zeta}x^2+2b\Im(\zeta)x+b^2}dx
   = a\int_{e^{-\lambda(T-t)}}^1\frac{b\Re(\zeta)i-\frac{1}{2}(2Ax+B)}{Ax^2+Bx+C}dx \nonumber \\
&= a\l[b\Re(\zeta)I(x)i-\frac{1}{2}\log|Ax^2+Bx+C|\r]_{e^{-\lambda(T-t)}}^1,
\label{eq-kappa-Gamma1}
\end{align}
where $A:=\zeta\overline{\zeta}$, $B:=2b\Im(\zeta)$, $C=b^2$ and
\[
I(x)=\frac{2}{\sqrt{4AC-B^2}}\arctan\l(\frac{2Ax+B}{\sqrt{4AC-B^2}}\r)=\frac{1}{b|\Re(\zeta)|}\arctan\l(\frac{\zeta\overline{\zeta}x+b\Im(\zeta)}{b|\Re(\zeta)|}\r).
\]
Thus, we have
\begin{align}
(\ref{eq-kappa-Gamma1})
&= ai\sgn(\Re(\zeta))\l\{\arctan\l(\frac{\zeta\overline{\zeta}+b\Im(\zeta)}{b|\Re(\zeta)|}\r)
   -\arctan\l(\frac{\zeta\overline{\zeta}e^{-\lambda(T-t)}+b\Im(\zeta)}{b|\Re(\zeta)|}\r)\r\} \nonumber \\
&  \hspace{5mm}-\frac{a}{2}\log\Big|\frac{\zeta\overline{\zeta}+2b\Im(\zeta)+b^2}{\zeta\overline{\zeta}e^{-2\lambda(T-t)}+2b\Im(\zeta)e^{-\lambda(T-t)}+b^2}\Big|.
\label{eq-kappa-Gamma2}
\end{align}
As a result, for $v\in\bbR$ and $\alpha\in(0,\widehat{u})$, substituting $v-i\alpha$ for $\zeta$ in (\ref{eq-kappa-Gamma2}), we obtain
\[
\l|\exp\l\{\int_t^T\kappa\l(i\zeta e^{-\lambda(T-s)}\r)ds\r\}\r| = \l|\frac{v^2+(\alpha-b)^2}{v^2e^{-2\lambda(T-t)}+(\alpha e^{-\lambda(T-t)}-b)^2}\r|^{-\frac{a}{2}},
\]
which is bounded on $v\in\bbR$.
Hence, (\ref{eq-phi}) implies that the gamma-OU case does not satisfy (\ref{eq-phicond}), but does (\ref{eq-prop-ass}) for any $a$, $b>0$.
\end{ex}

%
%
\setcounter{equation}{0}
\section{LRM strategies}
In this section, representations of the LRM strategies for the VIX call options are discussed.
A definition of the LRM strategies is given in Appendix \ref{A-LRM}.
Note that hedging strategies discussed in this paper are constructed by the underlying riskless and risky assets.
For any $t\in[0,T]$, we denote by $\xi^\calV_t$ the value of the LRM strategy at time $t$ for $(\calV_T-K)^+$ the VIX call option matured at time $T$ with strike price $K>0$.
In other words, an investor hedging the VIX call option $(\calV_T-K)^+$ in the LRM approach should hold $\xi^\calV_t$ units of the risky asset at time $t$.
On the other had, once $\xi^\calV_t$ is given, we can compute $\eta^\calV_t$ the amount of units of the riskless asset at time $t$ through (\ref{eq-eta}).
Thus, we give representations of $\xi^\calV_t$ alone.

Under the condition (\ref{eq-phicond}), a representation of $\xi^\calV_t$ is given as follows:

\begin{thm}\label{thm-LRM}
Suppose that (\ref{eq-phicond}) and
\begin{equation}\label{thm-LRM-cond}
\int_1^\infty\exp\l\{2\calB(T)x\r\}\nu(dx)<\infty,
\end{equation}
where $\calB(T)=\frac{1-e^{-\lambda T}}{\lambda}$.
Then, the LRM strategy $\xi^\calV$ exists, and, for any $t\in[0,T)$, $\xi^{\calV}_t$ is represented as follows:
\begin{align}\label{eq-thm-LRM}
\xi^\calV_t
&= \frac{e^{-r(T-t)}}{S_{t-}(\sigma^2_t+C_\rho)}\frac{1}{2\pi}\int_{-\infty}^\infty\widehat{g}(v,\alpha;K)\phi_{T|t-}(-v-i\alpha) \nonumber \\
&  \hspace{5mm}\times\int_0^\infty\l(\exp\l\{(-v-i\alpha)e^{-\lambda(T-t)}x\r\}-1\r)(e^{\rho x}-1)\nu(dx)dv,
\end{align}
where $\alpha\in(0,\widehat{u})$ and $C_\rho:=\int_0^\infty(e^{\rho x}-1)^2\nu(dx)$.
Note that the right-hand side of (\ref{eq-thm-LRM}) is independent of the choice of $\alpha$.
\end{thm}

\proof
This theorem is shown by Theorem A.1 of \cite{AIS-BNS} (see also Theorem 3.7 of \cite{AS}).
Thus, we confirm if all the conditions of Theorem A.1 of \cite{AIS-BNS} are satisfied in our setting.
Firstly, AS1 and AS2 are automatically satisfied, since $\widehat{S}$ is a martingale, and $(\calV_T-K)^+\in\bbD^{1,2}$ and $D_{t,0}(\calV_T-K)^+=0$ by Lemma \ref{lem-thm-LRM} below,
where $\bbD^{1,2}$ and $D_{t,0}$ are the Sobolev space and the Malliavin derivative operator respectively, defined in Appendix \ref{A-Malliavin}.
As for AS3, (\ref{eq-lem-thm-LRM}) below implies that
\[
\int_0^\infty(\bbE[xD_{t,x}(\calV_T-K)^+|\calF_{t-}])^2\nu(dx)\leq B_\calV \int_0^\infty x\nu(dx)
\]
for any $t\in[0,T]$.
Since $\int_0^\infty x\nu(dx)<\infty$ holds, we have
\[
\bbE\l[\int_0^T\int_0^\infty(\bbE[xD_{s,x}(\calV_T-K)^+|\calF_{s-}])^2\nu(dx)ds\r]<\infty,
\]
which implies the condition AS3.
In addition, we need to notice that Theorem A.1 of \cite{AIS-BNS} holds under their Assumption 2.2, which is satisfied in our setting under the condition (\ref{thm-LRM-cond}).
Remark that we can omit Item 2 of Assumption 2.2 of \cite{AIS-BNS}, since $\widehat{S}$ is a martingale.
Moreover, we do not need the finiteness of $\int_1^\infty\exp\l\{2|\rho|x\r\}\nu(dx)$, since it has been used to show AS2 in \cite{AIS-BNS}.
Consequently, Theorem A.1 of \cite{AIS-BNS} is available.

Theorem A.1 of \cite{AIS-BNS} implies that $\xi^\calV$ exists, and
\begin{equation}\label{eq-thm-LRM1}
\xi^\calV_t = \frac{e^{-r(T-t)}}{S_{t-}(\sigma^2_t+C_\rho)}\int_0^\infty\bbE[xD_{t,x}(\calV_T-K)^+|\calF_{t-}](e^{\rho x}-1)\nu(dx)
\end{equation}
holds, since $D_{t,0}(\calV_T-K)^+=0$ by Lemma \ref{lem-thm-LRM}.
Denoting
\[
\widetilde{\sigma}^2_{T|t}(x):=\sigma^2_t+xe^{-\lambda(T-t)}
\]
for $t\in[0,T]$ and $x\in(0,\infty)$, and using Lemma \ref{lem-thm-LRM} and (\ref{eq-VIX}), we can rewrite (\ref{eq-thm-LRM1}) as
\begin{align}\label{eq-thm-LRM2}
\xi^\calV_t
&= \frac{e^{-r(T-t)}}{S_{t-}(\sigma^2_t+C_\rho)}\int_0^\infty\Bigg(\bbE\l[\l(\sqrt{B_\calV\widetilde{\sigma}^2_{T|t}(x)+C_\calV}-K\r)^+\Big|\calF_{t-}\r] \nonumber \\
&  \hspace{5mm}-\bbE\l[\l(\sqrt{B_\calV\sigma^2_T+C_\calV}-K\r)^+\Big|\calF_{t-}\r]\Bigg)(e^{\rho x}-1)\nu(dx).
\end{align}
Remark that
\[
\bbE[\exp\{i\zeta\widetilde{\sigma}^2_{T|t}(x)\}|\calF_t]=\phi_{T|t}(\zeta)\exp\l\{i\zeta xe^{-\lambda(T-t)}\r\}.
\]
Thus, from the view of Proposition \ref{prop-Tankov}, denoting $\vt=-v-i\alpha$, we can rewrite (\ref{eq-thm-LRM2}) as
\begin{align*}
\xi^\calV_t
&= \frac{e^{-r(T-t)}}{S_{t-}(\sigma^2_t+C_\rho)}\int_0^\infty\frac{1}{2\pi}\int_{-\infty}^\infty\widehat{g}(v,\alpha;K)\phi_{T|t-}(\vt) \\
&  \hspace{5mm}\times\l(\exp\l\{i\vt e^{-\lambda(T-t)}x\r\}-1\r)dv(e^{\rho x}-1)\nu(dx) \\
&= \frac{e^{-r(T-t)}}{S_{t-}(\sigma^2_t+C_\rho)}\frac{1}{2\pi}\int_{-\infty}^\infty\widehat{g}(v,\alpha;K)\phi_{T|t-}(\vt) \\
&  \hspace{5mm}\times\int_0^\infty\l(\exp\l\{i\vt e^{-\lambda(T-t)}x\r\}-1\r)(e^{\rho x}-1)\nu(dx)dv
\end{align*}
for $\alpha\in(0,\widehat{u})$.
This completes the proof of Theorem \ref{thm-LRM}.
\fin

\begin{rem}
\begin{enumerate}
\item Under the condition (\ref{thm-LRM-cond}), it holds that $\widehat{u}>0$, since $\widehat{u}\geq2\calB(T)$.
\item As seen in \cite{AIS-BNS} and \cite{AS}, (\ref{thm-LRM-cond}) ensures the so-called (SC) condition, which is indispensable to discuss the LRM strategies.
\end{enumerate}
\end{rem} 

\begin{rem}
The IG-OU case satisfies all the conditions of Theorem \ref{thm-LRM} if $\frac{b^2}{2}>2\calB(T)$.
Thus, we can compute the LRM strategies $\xi^\calV_t$ for the IG-OU case through (\ref{eq-thm-LRM}) as long as $\frac{b^2}{2}>2\calB(T)$.
In addition, for $\zeta\in\bbC$ with $\Re(\zeta)<0$, we have
\begin{align*}
\int_0^\infty(e^{\zeta x}-1)(e^{\rho x}-1)\nu(dx) &= \kappa(\zeta+\rho)-\kappa(\zeta)-\kappa(\rho) \\
                                                  &= \lambda a\l\{\frac{\zeta+\rho}{\sqrt{b^2-2(\zeta+\rho)}}-\frac{\zeta}{\sqrt{b^2-2\zeta}}-\frac{\rho}{\sqrt{b^2-2\rho}}\r\}.
\end{align*}
\end{rem}

\begin{lem}\label{lem-thm-LRM}
For any $t\in[0,T]$ and $x\in[0,\infty)$, we have $(\calV_T-K)^+\in\bbD^{1,2}$ and
\begin{equation}\label{eq-lem-thm-LRM}
D_{t,x}(\calV_T-K)^+=\frac{1}{x}\l\{\l(\sqrt{B_\calV\widetilde{\sigma}^2_{T|t}(x)+C_\calV}-K\r)^+-(\calV_T-K)^+\r\}{\bf 1}_{\{x>0\}},
\end{equation}
where $\widetilde{\sigma}^2_{T|t}(x)=\sigma^2_T+xe^{-\lambda(T-t)}$.
Note that the definitions of the space $\bbD^{1,2}$ and the operator $D_{t,x}$ are given in Appendix \ref{A-Malliavin}.
\end{lem}

\proof
First of all, we show $\calV_T\in\bbD^{1,2}$; and calculate $D_{t,x}\calV_T$ for $t\in[0,T]$ and $x\in[0,\infty)$.
Now, we define $f(y)=\sqrt{B_\calV y+C_\calV}$ for $y\geq0$, that is, $\calV_T=f(\sigma^2_T)$.
Note that we can extend $f$ to a $C^1$-function on $\bbR$ with bounded derivative $f^\prime$.
Thus, since $\sigma^2_T\in\bbD^{1,2}$ and $D_{t,x}\sigma^2_T=e^{-\lambda(T-t)}{\bf 1}_{\{x>0\}}$ by Lemma A.2 of \cite{AIS-BNS},
Proposition 2.6 of \cite{Suz} implies that $\calV_T\in\bbD^{1,2}$,
\[
D_{t,0}\calV_T=f^\prime(\sigma^2_T)D_{t,0}\sigma^2_T=0,
\]
and
\[
D_{t,x}\calV_T=\frac{f(\sigma^2_T+xD_{t,x}\sigma^2_T)-f(\sigma^2_T)}{x}
              =\frac{\sqrt{B_\calV(\sigma^2_T+xe^{-\lambda(T-t)})+C_\calV}-\calV_T}{x}
\]
for $x\in(0,\infty)$.
Hence, Theorem 4.1 of \cite{AS} implies that
\[
D_{t,x}(\calV_T-K)^+=\frac{(\calV_T+xD_{t,x}\calV_T-K)^+-(\calV_T-K)^+}{x}{\bf 1}_{\{x>0\}},
\]
from which we obtain (\ref{eq-lem-thm-LRM}).
\fin

Next, we provide an approximate representation of $\xi^\calV_t$ under the condition (\ref{eq-prop-ass}) instead of (\ref{eq-phicond}).
This representation enables us to compute $\xi^\calV_t$ approximately for the gamma-OU case.

\begin{thm}\label{thm-LRM-ve}
Under the conditions (\ref{eq-prop-ass}) and (\ref{thm-LRM-cond}), $\xi^\calV$ exists, and we have
\begin{align}\label{eq-thm-LRM-ve}
\xi^\calV_t
&= \frac{e^{-r(T-t)}}{S_{t-}(\sigma^2_t+C_\rho)}\lim_{\ve\to0}\frac{1}{2\pi}\int_{-\infty}^\infty\widehat{g}(v,\alpha;K)\phi^{(\ve)}_{T|t-}(-v-i\alpha) \nonumber \\
&  \hspace{5mm}\times\int_0^\infty\l(\exp\l\{(-v-i\alpha)e^{-\lambda(T-t)}x\r\}-1\r)(e^{\rho x}-1)\nu(dx)dv
\end{align}
for any $t\in[0,T)$ and $\alpha\in(0,\widehat{u})$.
Note that the function $\phi^{(\ve)}_{T|t}$ is defined in (\ref{eq-phi-ve}), and the right-hand side of (\ref{eq-thm-LRM-ve}) is independent of the choice of $\alpha$.
\end{thm}

\proof
Denoting
\[
\widetilde{\sigma}^2_{T|t}(x,\ve) := \widetilde{\sigma}^2_{T|t}(x)+\ve(W_T-W_t) = \sigma^2_T+xe^{-\lambda(T-t)}+\ve(W_T-W_t),
\]
and
\[
A(u):=(\sqrt{|B_\calV u+C_\calV|}-K){\bf 1}_{\{u>(K^2-C_\calV)/B_\calV\}} \ \ \ \mbox{for any }u\in\bbR,
\]
we have
\begin{align*}
\lefteqn{\bbE\l[\l(\sqrt{B_\calV\widetilde{\sigma}^2_{T|t}(x)+C_\calV}-K\r)^+\Big|\calF_{t-}\r]} \\
&= \bbE[A(\widetilde{\sigma}^2_{T|t}(x))|\calF_{t-}] = \lim_{\ve\to0}\bbE[A(\widetilde{\sigma}^2_{T|t}(x,\ve))|\calF_{t-}].
\end{align*}
Thus, (\ref{eq-thm-LRM2}) and the dominated convergence theorem yield
\begin{align*}
\xi^\calV_t
&= \frac{e^{-r(T-t)}}{S_{t-}(\sigma^2_t+C_\rho)}\int_0^\infty\lim_{\ve\to0}\l\{\bbE[A(\widetilde{\sigma}^2_{T|t}(x,\ve))|\calF_{t-}]-\bbE[A(\sigma^2_{T|t}(\ve))|\calF_{t-}]\r\} \\
&  \hspace{5mm}\times(e^{\rho x}-1)\nu(dx) \\
&= \frac{e^{-r(T-t)}}{S_{t-}(\sigma^2_t+C_\rho)}\lim_{\ve\to0}\int_0^\infty\l\{\bbE[A(\widetilde{\sigma}^2_{T|t}(x,\ve))|\calF_{t-}]-\bbE[A(\sigma^2_{T|t}(\ve))|\calF_{t-}]\r\} \\
&  \hspace{5mm}\times(e^{\rho x}-1)\nu(dx),
\end{align*}
where $\sigma^2_{T|t}(\ve):=\sigma^2_T+\ve(W_T-W_t)$.
Now, denoting $\vt=-v-i\alpha$, we have
\begin{align*}
\lefteqn{\bbE[A(\widetilde{\sigma}^2_{T|t}(x,\ve))|\calF_{t-}]} \\
&= \frac{e^{-r(T-t)}}{2\pi}\int_{-\infty}^\infty\widehat{g}(v,\alpha;K)\phi_{T|t-}(\vt)
   \exp\l\{i\vt xe^{-\lambda(T-t)}-\frac{\ve^2\vt^2(T-t)}{2}\r\}dv
\end{align*}
for $\ve>0$ and $\alpha\in(0,\widehat{u})$ by the same sort of argument as Proposition \ref{prop-approx}.
As a result, Fubini's theorem implies
\begin{align*}
\xi^\calV_t
&= \frac{e^{-r(T-t)}}{S_{t-}(\sigma^2_t+C_\rho)}\lim_{\ve\to0}\int_0^\infty\frac{1}{2\pi}\int_{-\infty}^\infty\widehat{g}(v,\alpha;K)\phi_{T|t-}(\vt) \\
&  \hspace{5mm}\times\exp\l\{-\frac{\ve^2\vt^2(T-t)}{2}\r\}\l(\exp\l\{i\vt e^{-\lambda(T-t)}x\r\}-1\r)dv(e^{\rho x}-1)\nu(dx) \\
&= \frac{e^{-r(T-t)}}{S_{t-}(\sigma^2_t+C_\rho)}\lim_{\ve\to0}\frac{1}{2\pi}\int_{-\infty}^\infty\widehat{g}(v,\alpha;K)\phi_{T|t-}(\vt) \\
&  \hspace{5mm}\times\exp\l\{-\frac{\ve^2\vt^2(T-t)}{2}\r\}\int_0^\infty\l(\exp\l\{i\vt e^{-\lambda(T-t)}x\r\}-1\r)(e^{\rho x}-1)\nu(dx)dv
\end{align*}
for $\alpha\in(0,\widehat{u})$.
This completes the proof of Theorem \ref{thm-LRM-ve}.
\fin

\begin{rem}
For the gamma-OU case, (\ref{thm-LRM-cond}) is satisfied if
\begin{equation}\label{cond-gamma-OU}
b>2\calB(T).
\end{equation}
Note that the following is very useful when we compute (\ref{eq-thm-LRM-ve}) for the gamma-OU case:
\[
\int_0^\infty(e^{\zeta x}-1)(e^{\rho x}-1)\nu(dx) = ab\lambda\l(\frac{1}{b-\zeta-\rho}-\frac{1}{b-\zeta}-\frac{1}{b-\rho}+\frac{1}{b}\r)
\]
for $\zeta\in\bbC$ with $\Re(\zeta)<0$.
\end{rem}

%
%
\setcounter{equation}{0}
\section{Numerical experiments}
Our aim of this section is to compute the prices $P_t$ and the LRM strategies $\xi^\calV_t$ by using the FFT.
First of all, we introduce its basic idea by taking $P_t$ given in (\ref{eq-Pt}) as an example.
Defining a function $f$ as
\[
f(v):=\frac{\sqrt{B_\calV\pi}}{2(-iv+\alpha)^{\frac{3}{2}}}\erfc\l(K\sqrt{\frac{-iv+\alpha}{B_\calV}}\r)
\]
for $v\in\bbR$, we have
\[
\widehat{g}(v,\alpha;K)=\exp\l\{-\frac{(iv-\alpha)C_\calV}{B_\calV}\r\}f(v).
\]
In addition, defining the Fourier transform of $f$ as
\[
\widehat{f}(x):=\int_{-\infty}^\infty e^{-ivx}f(v)dv,
\]
we can rewrite (\ref{eq-Pt}) as
\[
P_t = \frac{e^{-r(T-t)}}{2\pi}e^{\frac{\alpha C_\calV}{B_\calV}}\widehat{f}\l(\frac{C_\calV}{B_\calV}\r).
\]
Thus, we can compute $P_t$ with the FFT.

\begin{rem}
Considering the vanilla option $(S_T-K)^+$ written on the underlying asset $S$, its price $P^V_t$ is given as
\[
P^V_t=\frac{e^{-r(T-t)}}{2\pi}\int_{-\infty}^\infty e^{(iv+1-\alpha)\log K}\frac{\phi_{T|t}(-v-i\alpha)\widehat{S}_{t-}^{-iv+\alpha}}{(v-i\alpha)(v-1-i\alpha)}dv
\]
by Example 2 of \cite{T}, where $\phi_{T|t}$ is the conditional characteristic function of $\widehat{S}_T$.
This is computable with a Fourier transform on $\log K$.
Hence, for computation on the VIX options, the FFT is used in a different way from the case of other options.
\end{rem}

Next, we implement numerical experiments on the prices $P_t$ and the LRM strategies $\xi^\calV_t$ for the gamma-OU case
by computing the right-hand sides of (\ref{eq-Pt-ve}) and (\ref{eq-thm-LRM-ve}) respectively with sufficient small $\ve>0$.
We use the parameter set estimated in \cite{Scho}, that is, we set $\rho=-1.2606$, $\lambda=0.5783$, $a=1.4338$, $b=11.6641$. Moreover, we fix $T=1$, $r=0.007$.
This parameter set satisfies the condition (\ref{cond-gamma-OU}).
Moreover, $\alpha$ is set to 1.75, and the observation period $\tau$ appeared in the definition of the VIX $\calV_T$ is fixed to $0.0833$, which is approximately one month.
In our numerical experiments, the asset price and the squared volatility at time $t$ are fixed to $S_t=1124.47$ and $\sigma_t^2=0.0145$ respectively, even if time $t$ may change.
Note that the values of the VIX at time $t$ is 0.18588.
We compute the right-hand sides of (\ref{eq-Pt-ve}) and (\ref{eq-thm-LRM-ve}) with $\ve=0.0001$ to obtain the values of $P_t$ and $\xi^\calV_t$ approximately.

\begin{figure}[h]
 \begin{minipage}{0.5\hsize}\vspace{-45mm}
    \hspace{-12mm}\includegraphics[width=100mm]{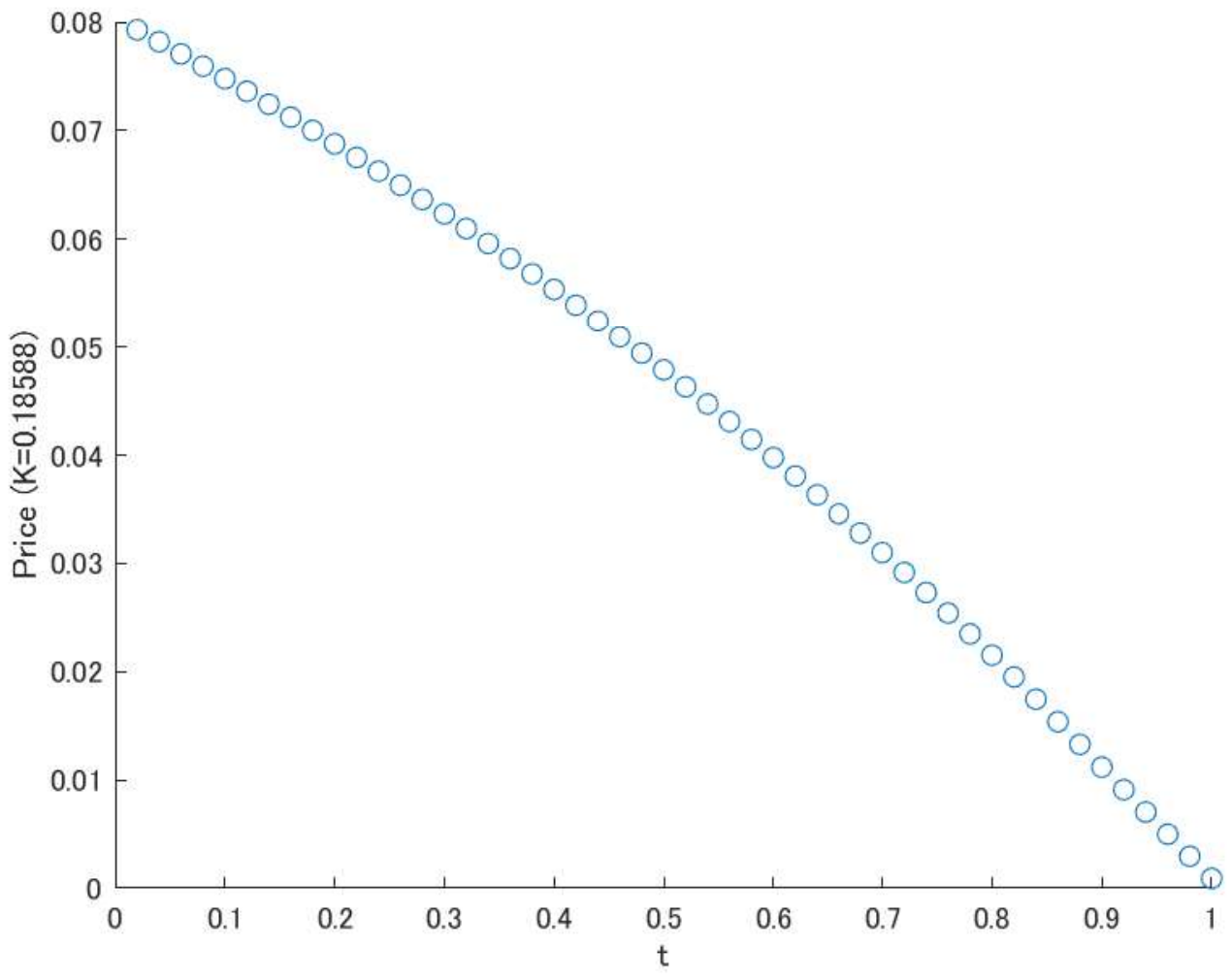}
    \vspace{-53mm}\captionsetup{width=.8\linewidth}\caption{Option prices versus times $t=0,0.02,\dots,0.98$ when the option is at the money.}\label{fig_price_t}
 \end{minipage}
 \begin{minipage}{0.5\hsize}\vspace{-48mm}
    \hspace{-17mm}\includegraphics[width=100mm]{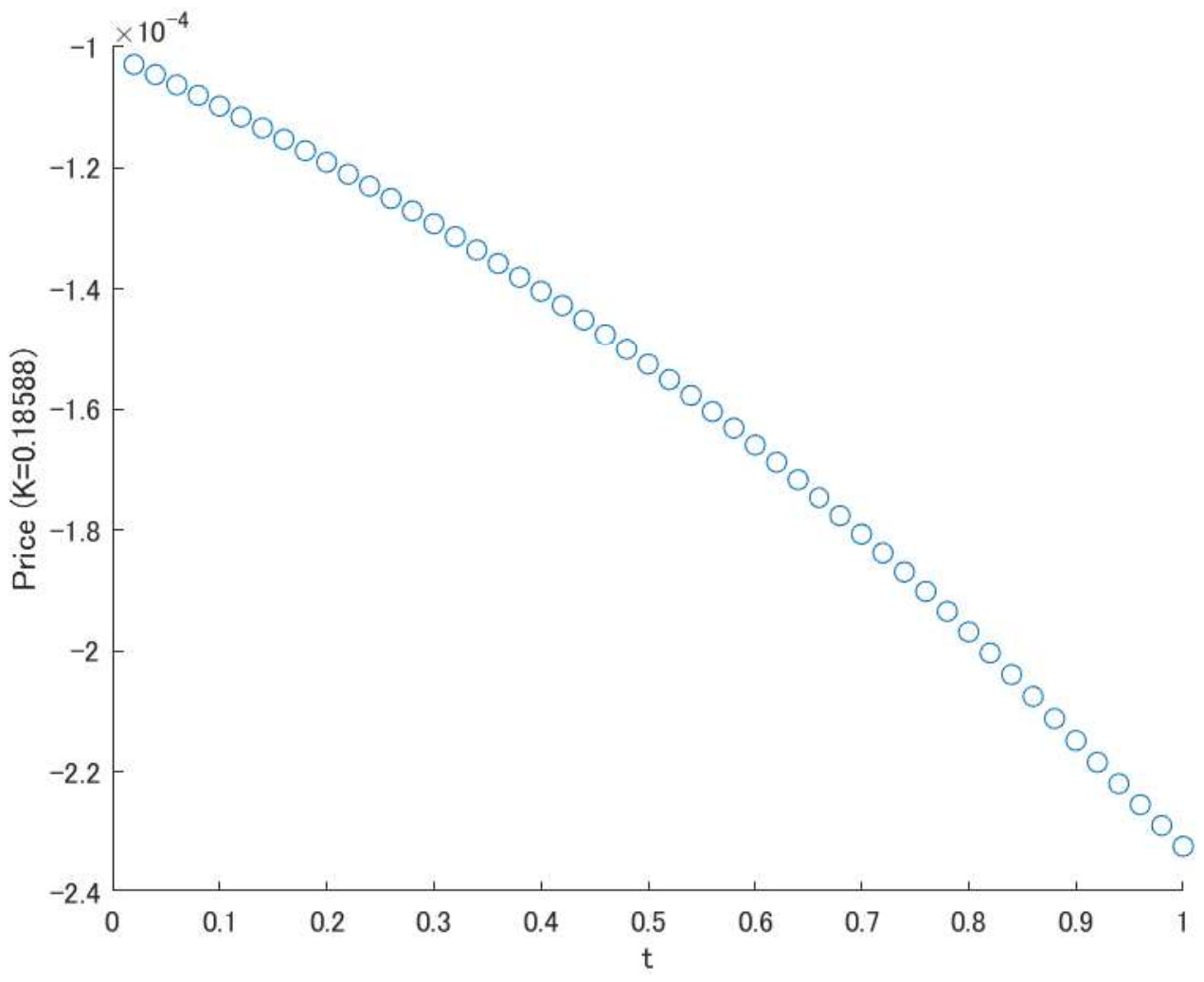}
    \vspace{-53mm}\captionsetup{width=.8\linewidth}\caption{Values of LRM strategies versus times $t$ for the at the money option.}\label{fig_LRM_t}
 \end{minipage}
 \begin{minipage}{0.5\hsize}\vspace{-45mm}
    \hspace{-12mm}\includegraphics[width=100mm]{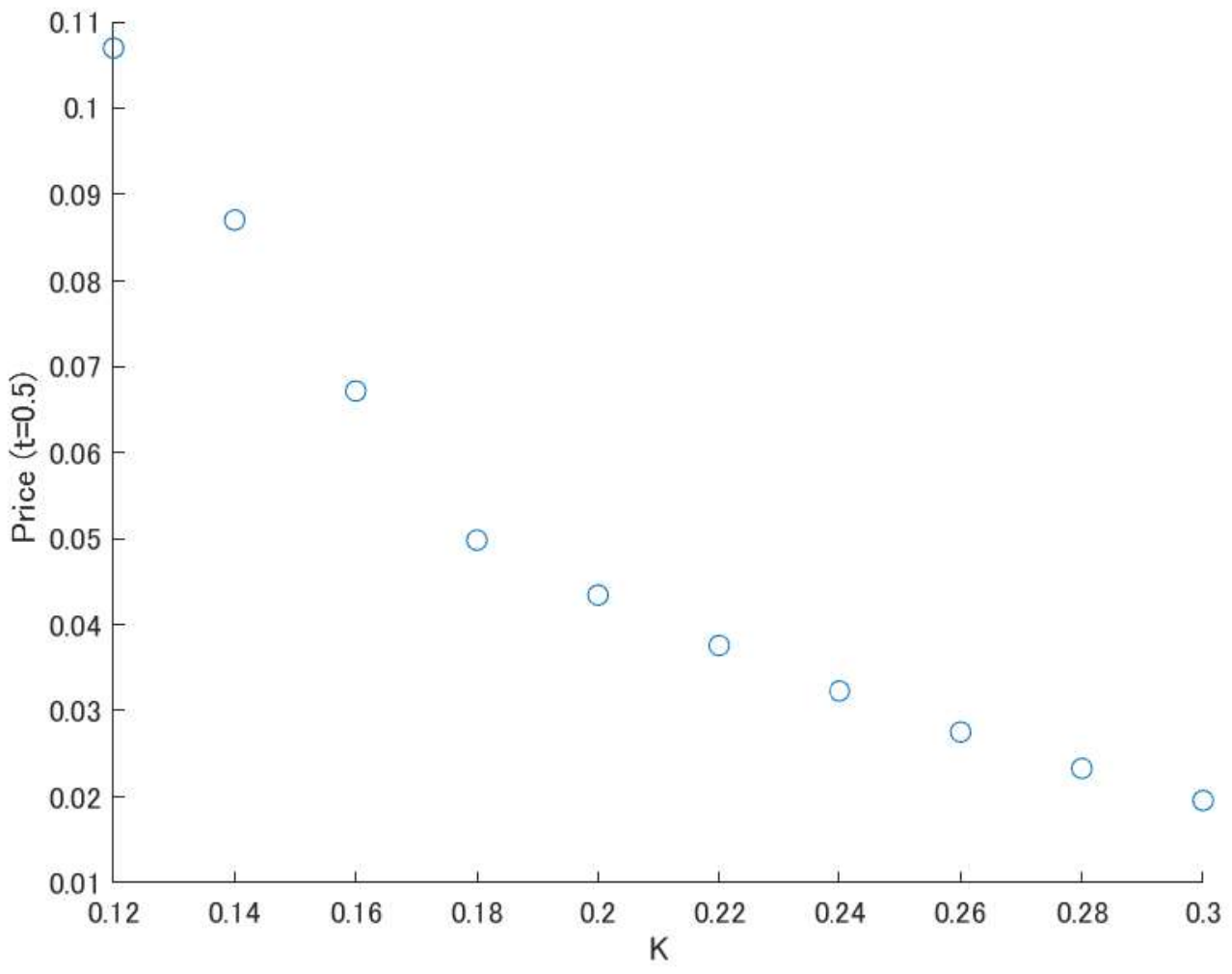}
    \vspace{-53mm}\captionsetup{width=.8\linewidth}\caption{Option prices at time $0.5$ versus strike prices $K$ from 0.12 to 0.3 at steps of 0.02.}\label{fig_price_K}
 \end{minipage}
 \begin{minipage}{0.5\hsize}\vspace{-48mm}
    \hspace{-17mm}\includegraphics[width=100mm]{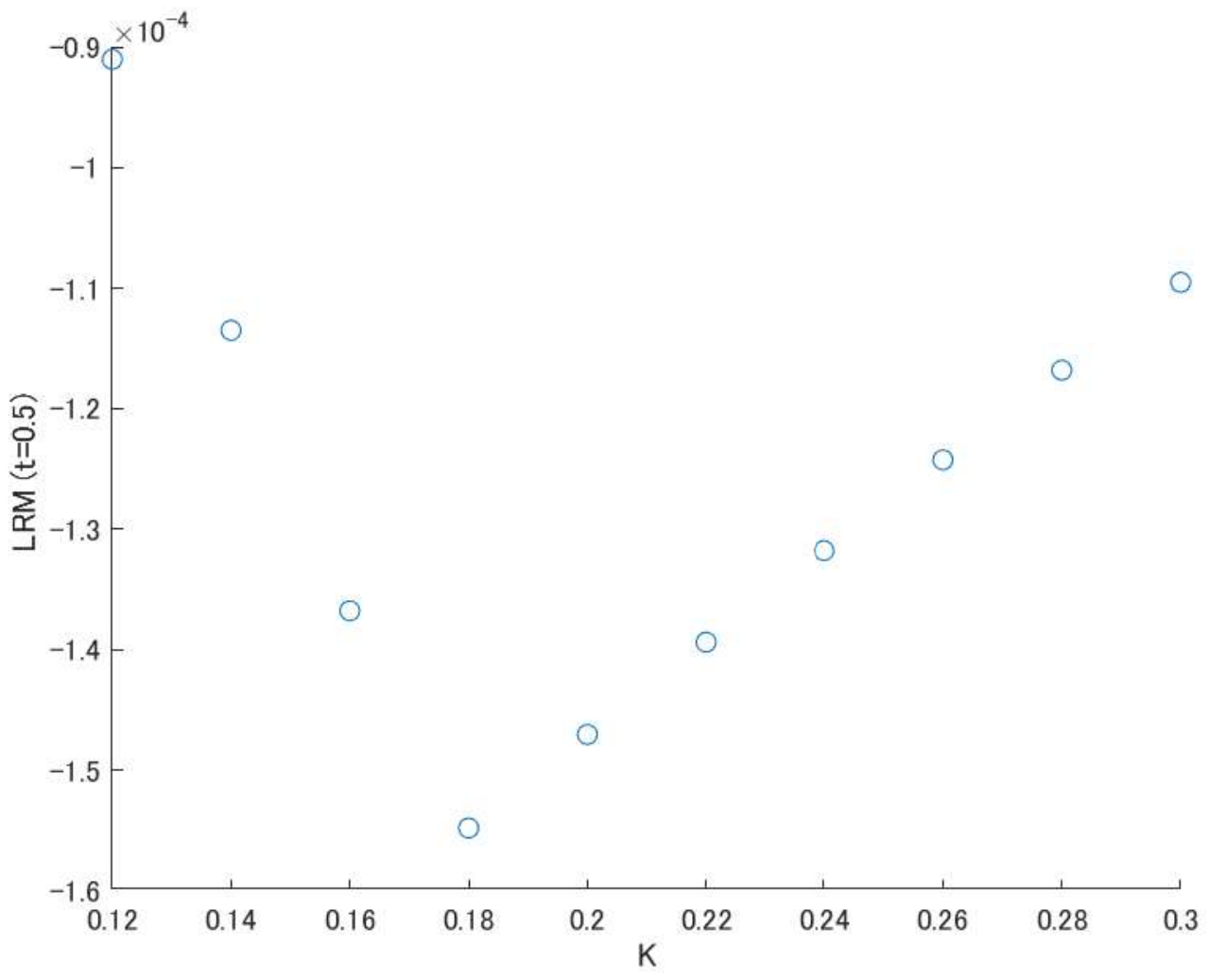}
    \vspace{-53mm}\captionsetup{width=.8\linewidth}\caption{Values of LRM strategies at time $0.5$ versus strike prices $K$.}\label{fig_LRM_K}
 \end{minipage}
\end{figure}

The following two types of experiments are implemented:
First, we compute the values of $P_t$ and $\xi^\calV_t$ for times $t=0, 0.02,\ldots,0.98$ when the option is at the money, that is, $K$ is fixed to 0.18588.
See Figures \ref{fig_price_t} and \ref{fig_LRM_t}.
Second, $t$ is fixed to 0.5, and we instead vary $K$ from 0.12 to 0.3 at steps of 0.02, and compute $P_t$ and $\xi^\calV_t$. See Figures \ref{fig_price_K} and \ref{fig_LRM_K}.
As seen in Figures \ref{fig_LRM_t} and \ref{fig_LRM_K}, the values of the LRM strategies are negative, since the VIX and the underlying risky asset have a negative correlation.

\appendix\normalsize
\setcounter{equation}{0}
\section{Appendix}
\renewcommand{\theequation}{A.\arabic{equation}}
\subsection{Malliavin calculus}\label{A-Malliavin}
We introduce Malliavin calculus for L\'evy processes briefly.
As stated in Remark \ref{rem-1}, we consider Malliavin calculus based on the canonical L\'evy space, undertaken by \cite{S07}.
For more details on this topic, see \cite{DI}, \cite{S07} and \cite{Suz}.

To begin with, we define two measures $q$ and $Q$ on $[0,T]\times[0,\infty)$ as
\[
q(E):=\int_E\delta_0(dx)dt+\int_Ex^2\nu(dx)dt,
\]
and
\[
Q(E):=\int_E\delta_0(dx)dW_t+\int_Ex\tN(dt,dx),
\]
where $E\in\calB([0,T]\times[0,\infty))$ and $\delta_0$ is the Dirac measure at $0$.
For $n\in\bbN$, we denote by $L_{T,q,n}^2$ the set of product measurable, deterministic functions $h:([0,T]\times[0,\infty))^n\to\bbR$ satisfying
\[
\|h\|_{ L_{T,q,n}^2}^2:=\int_{([0,T]\times[0,\infty))^n}|h((t_1,x_1),\cdots,(t_n,x_n))|^2q(dt_1,dx_1)\cdots q(dt_n,dx_n)<\infty.
\]
For $n\in\bbN$ and $h\in L_{T,q,n}^2$, we define
\[
I_n(h):=\int_{([0, T]\times[0,\infty))^n}h((t_1,x_1),\cdots,(t_n,x_n))Q(dt_1,dx_1)\cdots Q(dt_n,dx_n).
\]
Formally, we denote $L_{T,q,0}^2:=\bbR$ and $I_0(h):=h$ for $h\in\bbR$.
Under this setting, any square integrable $\calF_T$-measurable random variable $X$ has the unique representation
\[
X=\sum_{n=0}^{\infty}I_n(h_n)
\]
with functions $h_n\in L_{T,q,n}^2$ that are symmetric in the $n$ pairs $(t_i,x_i), 1\leq i\leq n$, and we have
\[
\bbE[X^2]=\sum_{n=0}^\infty n!\|h_n\|_{L_{T,q,n}^2}^2.
\]
We define the Sobolev space $\bbD^{1,2}$ and Malliavin derivative operator $D_{t,x}$ as follows:

\begin{defn}\label{def-Malliavin}
\begin{enumerate}
\item Let $\bbD^{1,2}$ denote the set of $\calF_T$-measurable random variables $X\in L^2(\bbP)$ with $X=\sum_{n=0}^\infty I_n(h_n)$ satisfying
      \[
      \sum_{n=1}^\infty nn!\|h_n\|_{L_{T,q,n}^2}^2<\infty.
      \]
\item For any $X\in\bbD^{1,2}$, the Malliavin derivative $DX:[0,T]\times[0,\infty)\times\Omega\to\bbR$ is defined as
      \[
      D_{t,x}X=\sum_{n=1}^\infty nI_{n-1}(h_n((t,x),\cdot))
      \]
      for $q$-a.e. $(t,x)\in[0,T]\times[0,\infty)$, $\bbP$-a.s.
\end{enumerate}
\end{defn}

\subsection{Definition of the LRM strategy}\label{A-LRM}
Before providing a definition of the LRM strategy, we prepare some terminologies.

\begin{defn}
\begin{enumerate}
\item A strategy is defined as a pair $\vp=(\xi, \eta)$, where $\xi$ is a predictable process, and $\eta$ is an adapted process.
      Note that $\xi_t$ and $\eta_t$ represent the amount of units of the risky and the riskless assets respectively which an investor holds at time $t$.
      The discounted value of the strategy $\vp=(\xi, \eta)$ at time $t\in[0,T]$ is defined as
      \[
      \widehat{V}_t(\vp):=\xi_t\widehat{S}_t+\eta_t.
      \]
      In particular, $\widehat{V}_0(\vp)$ gives the initial cost of $\vp$.
\item A strategy $\vp$ is said to be self-financing, if it satisfies
      \[
      \widehat{V}_t(\vp)=\widehat{V}_0(\vp)+\widehat{G}_t(\xi)
      \]
      for any $t\in[0,T]$,
      where $\widehat{G}(\xi)$ denotes the discounted gain process induced by $\xi$, that is,
      \[
      \widehat{G}_t(\xi):=\int_0^t\xi_sd\widehat{S}_s
      \]
      for $t\in[0,T]$.
      If a strategy $\vp$ is self-financing, then $\eta$ is automatically determined by $\xi$ and the initial cost $\widehat{V}_0(\vp)$.
\item For a strategy $\vp$, a process $\widehat{C}(\vp)$ defined by
      \[
      \widehat{C}_t(\vp):=\widehat{V}_t(\vp)-\widehat{G}_t(\xi)
      \]
      for $t\in[0,T]$ is called the discounted cost process of $\vp$.
      When $\vp$ is self-financing, its discounted cost process $\widehat{C}(\vp)$ is a constant.
\item Let $X$ be a square integrable random variable representing the payoff of a contingent claim at the maturity $T$.
      A strategy $\vp$ is said to replicate the claim $X$, if it satisfies $\widehat{V}_T(\vp)=\widehat{X}$,
      where $\widehat{X}:=e^{-rT}X$ the discounted value of $X$.
\end{enumerate}
\end{defn}

Finally, we give a definition of the LRM strategy $\vp^X$.
Roughly speaking, a strategy $\vp^X=(\xi^X,\eta^X)$, which is not necessarily self-financing, is called the LRM strategy for the claim $X$,
if it is the replicating strategy minimizing a risk caused by $\widehat{C}(\vp^X)$ in the $L^2$-sense among all replicating strategies.
The following definition is a simplified version based on Theorem 1.6 of Schweizer \cite{Sch3} under the assumption that $\widehat{S}$ is a martingale,
since the original one introduced by Schweizer \cite{Sch} and \cite{Sch3} is rather complicated.
Note that \cite{Sch3} treated the problem under the assumption that $r=0$. For the case where $r>0$, see, e.g. Biagini and Cretarola \cite{BC}.

\begin{defn}
\begin{enumerate}
\item A strategy $\vp=(\xi,\eta)$ is said to be an $L^2$-strategy, if $\xi$ is a predictable process satisfying
      \begin{equation}\label{eq-L2}
      \bbE\l[\int_0^T\xi^2_sd\la\widehat{S}\ra_s\r]<\infty,
      \end{equation}
      and $\eta$ is an adapted process such that $\widehat{V}(\vp)$ is a right continuous process with $\bbE[\widehat{V}_t^2(\vp)]<\infty$ for every $t\in[0,T]$.
\item An $L^2$-strategy $\vp$ is called the LRM strategy for the claim $X\in L^2(\bbP)$,
      if $\widehat{V}_T(\vp^X)=\widehat{X}$, and $[\widehat{C}(\vp^X),\widehat{S}]$ is a uniformly integrable martingale.
\item $X\in L^2(\bbP)$ admits a F\"ollmer-Schweizer decomposition, if it can be described by
      \[
      X=X_0+\int_0^T\xi^X_sd\widehat{S}_s+L_T^X,
      \]
      where $X_0\in\bbR$, $\xi^X$ is a predictable process satisfying (\ref{eq-L2}) and $L^X$ is a square-integrable martingale orthogonal to $\widehat{S}$ with $L_0^X=0$.
\end{enumerate}
\end{defn}

\noindent
Then, Proposition 5.2 of \cite{Sch3} or Proposition 3.7 of \cite{BC}, together with Remark 2.3 of \cite{AIS-BNS}, provides that, under the condition (\ref{thm-LRM-cond}),
the LRM strategy $\vp^X=(\xi^X,\eta^X)$ for $X\in L^2(\bbP)$ exists if and only if $\widehat{X}$($=e^{-rT}X$) admits a F\"ollmer-Schweizer decomposition
\[
\widehat{X}=\widehat{X}_0+\int_0^T\xi^{FS}_sd\widehat{S}_s+L_T^{FS},
\]
and its relationship is given by
\begin{equation}\label{eq-eta}
\xi^X_t=\xi^{FS}_t,\hspace{3mm}\eta^X_t=\widehat{X}_0+\int_0^t\xi^X_sd\widehat{S}_s+L^{FS}_t-\xi^X_t\widehat{S}_t.
\end{equation}
As a result, it suffices to obtain a representation of $\xi^X$ in order to get $\vp^X$.
Thus, we identify $\xi^X$ with $\vp^X$ in this paper.

\begin{center}
{\bf Acknowledgments}
\end{center}
The author gratefully acknowledges the financial support of the MEXT Grant in Aid for Scientific Research (C) No.15K04936 and No.18K03422.


\end{document}